\documentclass[12pt]{iopart}
\pdfoutput=1 
\usepackage[english]{babel}
\usepackage{iopams}
\usepackage{graphicx}
\usepackage[usenames,dvipsnames]{xcolor}
\usepackage{lipsum}
\usepackage{amssymb}
\usepackage{amsfonts}

\definecolor{orange}{RGB}{255,127,0}
\definecolor{maroon}{RGB}{128, 0, 0}
\definecolor{brown}{RGB}{150, 75, 0}

\usepackage[unicode=true,colorlinks=true,citecolor=BrickRed,urlcolor=BrickRed,linkcolor=BrickRed]{hyperref}

\graphicspath{{./figs/}}
\begin{document}

\title{\color{BlueViolet} Connecting neutrino physics with dark matter}  

\author{Massimiliano Lattanzi$^1$, Roberto A. Lineros$^2$, Marco Taoso$^3$}
\address{$^1$ Dipartimento di Fisica e Science della Terra, Universit\`a di Ferrara and INFN sezione di Ferrara, Polo Scientifico e Tecnologico - Edificio C Via Saragat, 1, I-44122 Ferrara, Italy\\}
\address{$^2$ Instituto de F\'{\i}sica Corpuscular -- CSIC/Universitad de Valencia, Parc Cient\'{\i}fic, calle Catedr\'{a}tico Jos\'{e} Beltr\'{a}n, 2, E-46980 Paterna, Spain\\}
\address{$^3$ Institut de Physique Th\'eorique, CNRS, URA 2306  CEA/Saclay,F-91191 Gif-sur-Yvette, France\\}

\date{\today}
\begin{abstract}
The origin of neutrino masses and the nature of dark matter are two of the most pressing open questions of the modern astro-particle physics. 
We consider here the possibility that these two problems are related, and review some theoretical scenarios which offer common solutions. 
A simple possibility is that the dark matter particle emerges in minimal realizations of the see-saw mechanism, like in the majoron and sterile neutrino
scenarios. 
We present the theoretical motivation for both models and discuss their phenomenology, confronting the predictions of these scenarios with cosmological and astrophysical observations. 
Finally, we discuss the possibility that the stability of dark matter originates from a flavour symmetry of the leptonic sector. We review a proposal based on an $A_4$ flavour symmetry.

\end{abstract}
\pacs{95.35.+d, 14.60.Lm, 14.60.St, 14.60.Pq, 11.30.Hv}
\submitto{\NJP}

\maketitle
%


\section{Introduction}
\label{sec:intro}

It is by now a well-established fact that a wide variety of cosmological observations 
provide strong support for the $\Lambda$CDM model, that can thus be regarded as
the Standard Cosmological Model (SCM). These observations include measurements of the 
temperature and polarization anisotropies of the cosmic
microwave background (CMB)~\cite{Ade:2013kta,Ade:2013ktc,Ade:2013zuv,Bennett:2012zja,Hinshaw:2012aka,Das:2013zf,Sievers:2013ica,Story:2012wx}, of the distribution
of large scale structures~\cite{Anderson:2013zyy,Parkinson:2012vd,Reid:2009xm}, 
of the abundance of light elements~\cite{Cooke:2013cba,Iocco:2008va,Izotov:2013waa},
 of the present value of the Hubble constant
$H_0$~\cite{Freedman:2012ny,Riess:2011yx}, 
of the magnitude-redshift relationship for Type Ia supernovae~\cite{Conley:2011ku,Suzuki:2011hu}.
According to the SCM, the Universe is spatially flat and its 
present energy density is dominated by non-relativistic matter (roughly 30\% of the total), in the form of baryons and cold dark matter (CDM), 
and dark energy (the remaining 70\%), in the form of a cosmological constant, that is
responsible for the present accelerated expansion. Photons and neutrinos are subdominant today
but their energy density drove the expansion of the Universe during the early radiation-dominated phase.
The structures that we observe today have grown from primordial adiabatic, nearly scale-invariant
fluctuations generated after a phase of inflationary expansion, in the simplest models driven
by the energy density of a scalar field~\cite{1981PhRvD..23..347G,1982PhLB..108..389L}.

However, in spite of its phenomenological success, one of the striking features of the 
SCM, from a theoretical point of view, is that we do not know the nature of nearly 95\% of the total matter-energy
content of the Universe.
In particular, only a part of the total matter density can be provided
by SM particles. The density of baryons can be inferred in several ways, mainly through the knowledge
of light elements abundances (that are extremely sensitive to the baryon-to-photon ratio) and from
the CMB anisotropies (since the presence of baryons induces a peculiar alternating pattern in the peaks
of the power spectrum). On the other hand the total density of matter affects,  
through gravity, both the average expansion and the evolution of perturbations, and thus
can be constrained, among others, by measuring the clustering properties of galaxies, or again 
by CMB observations. All the observational evidence points towards a coherent picture where
baryons constitute roughly 20\% of the total cosmological matter content, while the remaining 80\%
is provided by an electromagnetically neutral component dubbed dark matter. Since 
the gravitational evidence for dark matter comes from observations at different scales, it is difficult
to explain these anomalies in terms of a modified theory of gravity.

Dark matter also drives the process of formation of cosmological structures, as it creates the potential
wells were the luminous matter - i.e. baryons - falls once it is free from the support of photon pressure. In fact, the observed
clustering properties of galaxies also constrain the velocity dispersion of the dark matter component, since
this defines the free-streaming length below which perturbations in the dark matter density are erased and
clustering is suppressed.
This rules out ``hot'' dark matter candidates (HDM), like the SM neutrinos themselves,
whose large velocity dispersion results in a cut off in the matter power spectrum well above the galactic scale
(i.e., well above a few comoving Mpcs). Thus dark matter has to be ``cold'' (CDM) or ``warm'' (WDM), namely, with 
a damping length below or around the galactic scale, respectively. The predictions
of the CDM and WDM scenarios at the largest scales are identical, however WDM
has been often invoked as a possible solution for the shortcomings of CDM at small scales,
like those related to the abundance of dwarf satellites and to the inner density profile of galaxies. On the other
hand, complex baryonic physics could also be responsible for this, and this issue is still 
matter of intense debate in the structure formation community.

The exact nature of dark matter is still a mystery to date. Many candidates have been proposed for the role of dark matter, 
with different particle physics motivation.
Compiling a comprehensive list would be impossible, however popular examples include the supersymmetric particles, 
Kaluza-Klein particles, the axion. Unfortunately, the search for these or any other candidate has been unfruitful so far, and all theoretical possibilities are still open. 
In this review we will focus on the theoretically appealing possibility  that dark matter is somewhat related to neutrinos, and in particular to the mechanism
of neutrino mass generation, which is another open question of theoretical physics.

We know from the observation of neutrino oscillations that neutrinos have masses~\cite{Tortola:2012te,Capozzi:2013csa}; 
however their origin and nature (Dirac or Majorana) is still unknown. Moreover, the smallness of neutrino masses with respect to those of the SM charged fermions, remains a puzzle.
An elegant solution to these issues is provided by the see-saw mechanism~\cite{minkowski:1977sc,gellmann:1979,Yanagida::1979,Lazarides:1980nt,Mohapatra:1979ia,Schechter:1980gr}, described in Sec.~\ref{sec:mass_gen},
that allows to generate neutrino masses that are naturally smaller than those of the other fermions.
In this scenario neutrinos are usually Majorana particles, implying that the lepton number is violated. Therefore,
searches of lepton-number violating processes, in particular neutrinoless double $\beta$ decay, are a decisive test for unveiling the nature of neutrinos.

The general idea behind the see-saw mechanism can be embedded in many distinct scenarios, that often provide in turn viable dark matter candidates. Here we describe two simple and direct connections between the neutrino mass generation  and dark matter.
In Sec.~\ref{sec:majoron} we focus on the possibility that dark matter is the  Goldstone boson associated to the spontaneous breaking of the ungauged lepton number. This particle, called Majoron, can acquire a mass through quantum gravity effects that explicitly break global symmetries, and thus play the role of  dark matter.
In the case of sterile neutrinos dark matter, that we review in Sec.~\ref{sec:sterile}, three right-handed singlets are added to the particle content of the SM; 
dark matter is made from the lightest sterile state, while the two heaviest states generate the active neutrino masses through the see-saw mechanism.
In these two scenarios, the mass of the dark matter particle is not predicted by the theory; however, the keV mass range is favoured 
by observations, especially in the case of the sterile neutrino.

The majoron and sterile neutrino scenarios that we are considering, are both based on a minimal implementation of the seesaw mechanism.
Dark matter candidates can also arise when the seesaw is embedded in more complex models, for instance those in which neutrino
masses are generated radiatively (e.g.~\cite{2006PhRvD..73g7301M,2013JHEP...10..149H,2013PhRvL.110u1802G,Restrepo:2013aga,Krauss:2002px,Law:2013saa,Aoki:2014cja}), in R-parity violating SUSY models \cite{Hirsch:2005ag,Restrepo:2011rj,Diaz:2011pc}, or in different types of low-scale seesaw scenarios (e.g.~\cite{Fabbrichesi:2014qca,JosseMichaux:2011ba,Arina:2008bb,DeRomeri:2012qd,Basso:2012ti,An:2011uq,Lindner:2014oea}). The resulting candidates can lie at different mass scales and therefore exhibit a completely distinct phenomenology with respect to the sterile neutrino and the majoron.

We give a concrete example, in Sec.~\ref{sec:flavour}, considering a scenario where the dark matter candidate is not only related to the generation of neutrino masses but also to the presence of a discrete flavour symmetry, which at the same time approximately accounts for the observed pattern
of neutrino oscillations and stabilizes the dark matter particles.

\section{Generation of neutrino masses}
\label{sec:mass_gen}

The neutrino masses can be generated at non-renormalizable level by the effective dimension-five operator~\cite{Weinberg:1980bf}:
\begin{equation}
  \mathcal{O}_5 \propto (L_i H)^T (L_j H) \, , \label{eq:dim5}
\end{equation}
where $L_i$ denotes an electroweak doublet ($i,\, j = e,\,\mu,\,\tau$) and $H$ is 
the SM higgs doublet. This operator violates lepton number by two units and, 
after the electroweak symmetry is broken and the higgs 
acquires a vacuum expectation value (vev) $v_2\equiv\langle H \rangle$, generates
Majorana neutrino masses $m_\nu\propto v_2^2$. Assuming that $\mathcal O_5$
is generated at tree level by the exchange of some heavy degree of freedom with mass $M$,
one has 
\begin{equation}
m_\nu = \lambda\frac{v_2^2}{M} \simeq \frac{m_D^2}{M},
\end{equation}
where $\lambda$ is a dimensionless coupling, and the last equality follows from the fact 
that $v_2$ sets the mass scale $m_D$ of the charged fermions. The mass of the heavy particles
corresponds to the energy scale at which lepton number is violated. This can be pushed arbitrarily high,
in order to obtain $m_\nu \ll m_D$ and thus explain in a natural way the smallness of neutrino masses with respect
to the other SM fermions. If lepton number violation is related
to some gravitational Planck-scale effect we expect $M\simeq M_{\mathrm{Pl}}$, but the resulting
neutrino mass (assuming $\lambda \simeq 1 $) would be too small to explain the observed
mass differences. This suggests that neutrino masses are generated by new physics below the Planck scale.

\subsection{Seesaw models}
In seesaw models \cite{minkowski:1977sc,gellmann:1979,Yanagida::1979,Lazarides:1980nt,Mohapatra:1979ia,Schechter:1980gr},
 the effective operator (\ref{eq:dim5}) is generated at tree level
by the exchange of heavy particles. This can be achieved in different ways. A first possibility
is to add $\mathrm{SU(2)_L \otimes U(1)_Y}$ singlet
right-handed fermions (``sterile neutrinos'') ${\nu_R}$ to the particle 
content of the SM\footnote{The number of generations of these singlet neutrinos
is somewhat arbitrary, but for aesthetic reasons we will take it to be three, unless otherwise stated.}.
This allows to introduce the following term in the Yukawa lagrangian:
\begin{equation}
  \mathcal{L}_2 = Y_2 \, \bar{L} H {\nu_R} + \mathrm{h.c.} \, , \label{eq:diracmass}
\end{equation}
that, after electroweak symmetry breaking, generates  
a Dirac mass term for neutrinos, with $m_D = Y_2 v_2$. Moreover,
because the ${\nu_R}$ are gauge singlets, we can also add to the lagrangian
a gauge-invariant bare mass term for the right-handed neutrinos:
\begin{equation}
\mathcal{L}_R = \frac{1}{2} M_R \, {\nu_R}^{T}  {\nu_R} + \mathrm{h.c.} \, . \label{eq:maj-rh}
\end{equation}
This term can also be generated by interactions with a singlet scalar fiels, as 
in majoron models (see below).
A similar bare mass term for the active neutrinos is forbidden
as it would spoil the gauge invariance of the SM. Thus, the neutrino mass matrix in the $(\nu,\,{\nu_R}^{c})$ basis reads:
\begin{equation}
\label{eq:mass_matrix}
\mathcal{M}_\nu = \left(
\begin{array}{cc}
0 &  m_D \\
m_D & M_R
\end{array}
\right) \, .
\end{equation}
 If the Majorana mass $M_R$ (that is related to some physics beyond the SM)
is much larger than the Dirac mass $ m_D \equiv Y_2 v_2 $, 
then the mass eigenvalues are approximately $M_R$ and $m_D^2/M_R\ll M_R$. This is
the \emph{type-I} seesaw \cite{minkowski:1977sc,gellmann:1979,Yanagida::1979,Mohapatra:1979ia}.

In the \emph{type-II} seesaw~\cite{Mohapatra:1979ia,Magg:1980ut,Lazarides:1980nt,Schechter:1980gr,Schechter:1981cv}, instead, 
a scalar Higgs triplet $\Delta$ is added to the theory,
schematically coupling to the ordinary neutrinos through
\begin{equation}
\mathcal{L}_3 = {Y_3} \, L^{T} \Delta L + \mathrm{h.c.} \, . \label{eq:yuk3}
\end{equation}
Once the triplet acquires a vev $v_3\equiv \langle\Delta\rangle$, it gives rise 
to a Majorana mass term for active neutrinos, with $M_L = Y_3 v_3$. 
The triplet vev induces a change in the electroweak parameter $\rho$ \cite{Agashe:2014kda}, so it
is experimentally constrained to lie below a few GeVs \cite{Gunion:1989ci,Blank:1997qa}. 
This experimental evidence implies $v_3 \ll v_2$.
It is worth stressing that one can consider a more general seesaw model where both the right-handed
singlets and the scalar triplet are present, so that the neutrino mass matrix has the full seesaw structure:
\begin{equation}
\mathcal{M}_\nu = \left(
\begin{array}{cc}
M_L &  m_D \\
m_D & M_R
\end{array}
\right) \, .
\end{equation}
with $M_L \ll m_D \ll M_R$. We also note that the majorana mass terms
violate lepton number conservation.\\

Another possibility to generate the effective operator  \ref{eq:dim5} is through the tree-level exchange
of $\mathrm{SU(2)_L}$ fermion triplets $\Sigma$, as in the \emph{type-III} seesaw \cite{Foot:1988aq}.

\section{Majoron dark matter}
\label{sec:majoron}
\subsection{The model}

The basic idea behind the majoron model, originally proposed by Chikashige, Mohapatra and Peccei \cite{chikashige:1981ui} in 1980,
is that the global lepton number symmetry of the standard model of particle physics is spontaneously broken.
Indeed, as noted above, if neutrinos are majorana particles, lepton number is necessarily broken. If lepton number is
a global symmetry, then after its breakdown a massless Nambu-Goldstone boson - the majoron - is generated.

In the simplest version of the majoron model, three singlet neutrinos ${\nu_R}$ are added to the SM, allowing
to generate a Dirac mass for the neutrinos through a term of the form 
(\ref{eq:diracmass}). A complex
scalar higgs singlet $\sigma$ with lepton number 2 is also introduced, coupling to the singlet neutrinos through:
\begin{equation}
\mathcal{L}_1 = \frac{1}{2} Y_1 \, {\nu_R}^{T} \sigma {\nu_R} + \mathrm{h.c.} \, .
\end{equation}
As we shall see, $\sigma$ is the parent field of the majoron.
For the moment, we just note that when $\sigma$ acquires a vev $v_1\equiv\langle\sigma\rangle$, lepton number is broken and
 a Majorana mass term for the singlet neutrinos is generated, like the one in Eq. (\ref{eq:maj-rh}), with $M_R
\equiv Y_1 v_1$. \\
Thus, the neutrino mass matrix in the $(\nu,\,{\nu_R}^c)$ basis has the type-I seesaw structure:
\begin{equation}
\mathcal{M}_\nu = \left(
\begin{array}{cc}
0 &  v_2 Y_2 \\
v_2 Y_2 & v_1 Y_1
\end{array}
\right) \, ,
\end{equation}
where the condition $v_2\ll v_1$ is required in order to ensure the smallness of neutrino masses.
In the simplest implementation of the model, the absence of a higgs triplet implies that
there is no Majorana mass term for the ordinary neutrinos.

 Once $\sigma$ has acquired a vev (and lepton number is broken), we can write
 \begin{equation}
 \sigma = \frac{1}{\sqrt{2}}\Big( \langle\sigma \rangle + \rho + iJ \Big)
 \end{equation}
 where $\rho$ and $J$ are, respectively, a massive and massless boson field with zero vevs.
 The field $J$ is the majoron, the Goldstone boson associated to spontaneous breaking of lepton number.
 
 The model was generalized shortly after by Schechter \& Valle \cite{Schechter:1981cv} by 
 allowing for the presence of all Higgs multiplets - singlet, doublet and triplet. 
 Thus, with respect to the model sketched above, a Higgs triplet $\Delta$ and
 the corresponding term (\ref{eq:yuk3}) in the Yukawa lagrangian are also introduced.
 Once the triplet acquires a vev $v_3$, a Majorana mass term
 for the active neutrinos is generated. The full neutrino mass matrix has thus the 
 general seesaw structure:
 \begin{equation}
\mathcal{M}_\nu = \left(
\begin{array}{cc}
v_3 Y_3 &  v_2 Y_2 \\
v_2 Y^T_2 & v_1 Y_1
\end{array}
\right) \, ,
\end{equation}
where this time, in writing the matrix, we have also taken into account that the Yukawa couplings
are actually $3\times 3$ matrices. The three vevs satisfy $v_1 \gg v_2 \gg v_3$ as well as a 
vev seesaw relation $v_1\,v_3 \simeq v_2^2$. The resulting light neutrino mass is
\begin{equation}
m_\nu \simeq Y_3\,v_3 - Y_2\, Y_1^{-1} \,Y_2^T\, \frac{v_2^2}{v_1} \, .
\end{equation}
The properties of the majoron in this more general scenario can be derived using the invariance of the 
potential under lepton number and weak hypercharge symmetries \cite{Schechter:1981cv}. In particular, the majoron is given, apart from a normalization factor,
 by the following combination of the Higgs fields:
\begin{equation}
J \propto  v_3 {v_2}^2 \,\mathrm{Im}(\Delta^0) - 2 v_2 {v_3}^2 \,\mathrm{Im}(\Phi^0) 
+ v_1 ( {v_2}^2 + 4 {v_3}^2) \, \mathrm{Im}(\sigma) \, ,
\end{equation}
where $\Phi_0$ and $\Delta_0$ are the neutral components of the doublet and triplet, respectively.
It is clear that the pure singlet model can be recovered by setting $v_3 = 0$. 
The hierarchy of the vevs anyway implies that the majoron has to be dominantly singlet.
Again using the symmetry properties one can show that the majoron couples to the light neutrinos
proportional to their mass and inversely proportional to $v_1$ \cite{Schechter:1981cv} . 

The majoron, being a Goldstone boson, is massless.
However, it has been conjectured that if gravity violates 
global symmetries (see e.g. Ref.~\cite{coleman:1988tj}), then the majoron may acquire a mass through nonpertubative gravitational 
effects~\cite{akhmedov:1992hi}.
The value of its mass $m_J$ cannot be unambiguously calculated, as it depends on the details of how this explicit breaking of global symmetries occurs, which are rather unknown.

A massive majoron can decay at tree level to a pair of light neutrinos with a rate given by~\cite{akhmedov:1992hi} 
\begin{equation}
  \label{eq:Jtonu}
\Gamma_{J\rightarrow\nu\nu} = \frac{m_J}{32\pi} \frac{\sum_i (m^\nu_i)^2}{2 v_1^2} \, .
\end{equation}
Moreover, the majoron also possesses a subleading decay mode to two photons, induced at the loop level
through its coupling to the charged fermions. The decay rate for this process is (in the limit $m_J \ll m_f$)~\cite{bazzocchi:2008fh}
\begin{equation}
\label{eqJ:gg}
  \Gamma_{J\rightarrow\gamma\gamma} = \frac{\alpha^2 m_J^3}{64\pi^3}  \left| \sum_f N_f  Q_f^2
\frac{2 v_3^2}{v_2^2 v_1} (- 2 T_{3}^f)
    \frac{m_J^2}{12 m_f^2}\right|^2\,,
\end{equation}
where $N_f$, $Q_f$, $T_3^f$ and $m_f$ denote respectively the color
factor, electric charge, weak isospin and mass of the SM electrically
charged fermions $f$. The radiative decay rate is proportional to the triplet
vev $v_3$ and is thus a peculiar feature of the more general seesaw model.

\subsection{Majoron cosmology and astrophysics}
A massive majoron has many astrophysical and cosmological implications that were
first explored in Refs.~\cite{akhmedov:1992hi,Rothstein:1992rh,Cline:1993ht,berezinsky:1993fm}.
In particular, in Refs. \cite{Rothstein:1992rh,berezinsky:1993fm} it was suggested that
the majoron could play be the role of the dark matter particle. The majoron could be produced
in the early Universe either thermally or through some non thermal mechanism,
like for example, a phase transition \cite{berezinsky:1993fm} or the evaporation of majoron strings
  \cite{Rothstein:1992rh}.

In order for the majoron to be the dark matter, however, its lifetime must be large enough 
for it to be stable on cosmological timescales. CMB data can been used 
to constrain the majoron decay rate to neutrinos \cite{Lattanzi:2007ux}. 
An analysis of the WMAP 9-year data \cite{Bennett:2012zja} yields
\cite{Lattanzi:2013uza}
\begin{equation}
\Gamma_{J\to\nu\nu} \le 6.4\times 10^{-19}\,\mathrm{s}^{-1} \qquad (95\%\;\textrm{C.L.})\, ,
\label{eq:gammanu}
\end{equation}
corresponding to $\tau_J \ge 50\,\mathrm{Gyr}$. Very recently, a combined analysis of the Planck 2013, WMAP polarization,
WiggleZ and BOSS data has allowed to improve this limit to \cite{Audren:2014bca}
\begin{equation}
\Gamma_{J\to\nu\nu} \le 1.9\times 10^{-19}\,\mathrm{s}^{-1} \qquad (95\%\;\textrm{C.L.})\, , 
\label{eq:gammanu2}
\end{equation}
or $\tau \ge 160\,\mathrm{Gyr}$. 
In these analysis the cosmological data are fitted assuming that majorons make up for all the dark matter content of the Universe. The majoron decay rate is an extra parameter with respect to the six parameters of the standard $\Lambda$CDM model. As expected, the obtained majoron energy density is very close to the dark matter density derived for the standard $\Lambda$CDM model. In fact, the decay rate is constrained to be very small so that the majoron effectively acts as a stable dark matter particle on cosmological timescales.

If the majoron was in thermal equilibrium in the early Universe 
with the SM plasma, and decoupled while still relativistic, the inferred relic density
singles out the sub-keV range for the majoron mass. The keV mass range is, in principle, very interesting 
as it would make the majoron, assuming again it is a thermal relic,
 a WDM candidate and could help to solve the problems of the standard CDM paradigm
at the galactic scales; however, a mass in the sub-keV range is potentially problematic as well
as it would probably lead, in the case of a thermal spectrum, to an excessive cancellation of small-scale perturbations.
However, larger majoron masses are possible in the case of non-thermal production mechanisms.
\begin{figure}[tb]
	\centering
	\includegraphics[width=0.5\columnwidth]{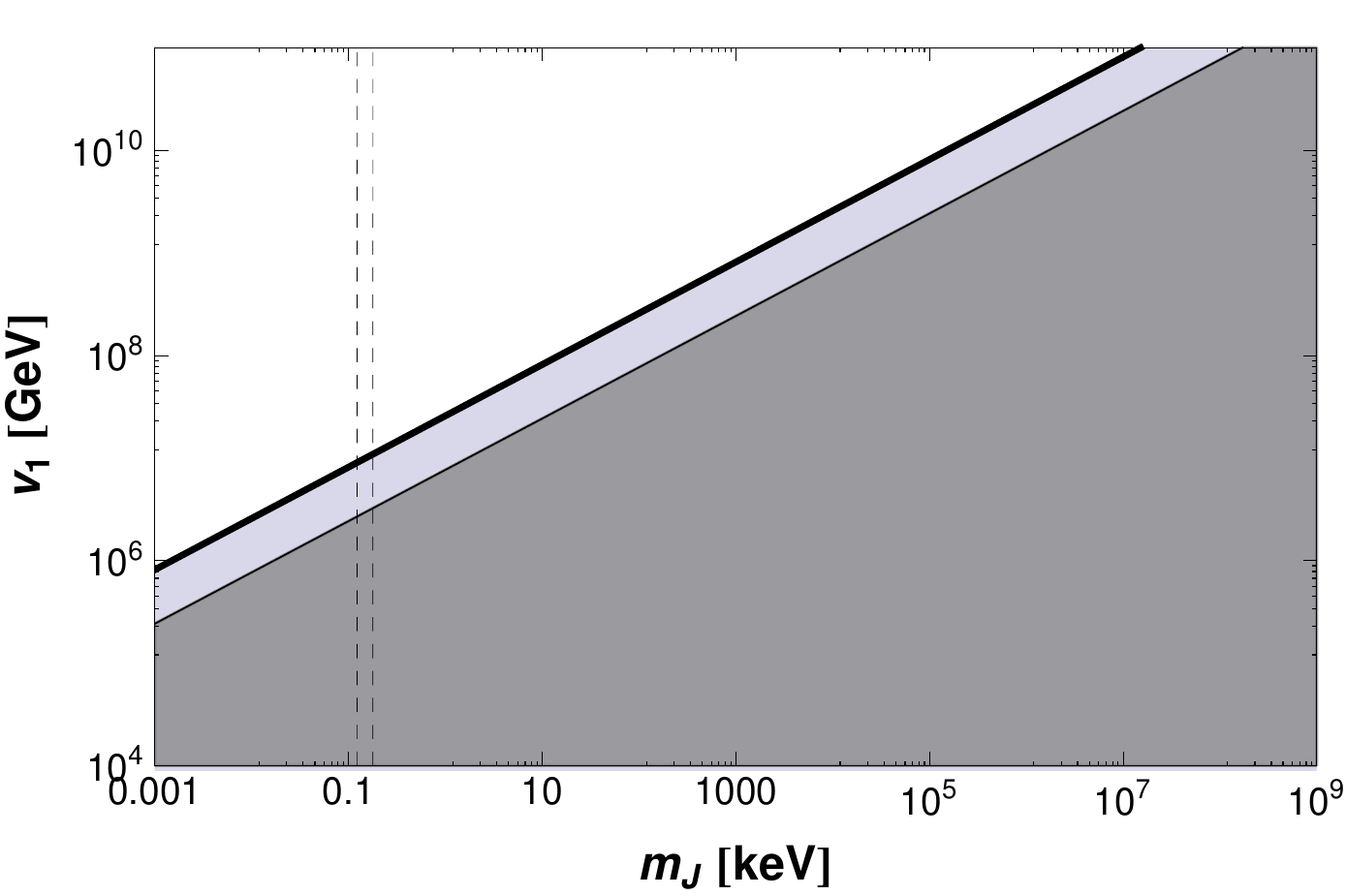}
	\caption{Constraints on the Majoron vev $v_1$ and mass $m_J$, see Eq. (\ref{eq:v1bound}). The shaded region
	below the lines are excluded by cosmological observations as the Majoron would decay too fast. The thick (thin) line corresponds
	to the upper limit on the decay rate coming from Planck 2013, WMAP9 polarization, WiggleZ and BOSS data (WMAP9 data) \cite{Lattanzi:2013uza,Audren:2014bca}. 
	We have assumed normal hierarchy and
	a vanishing mass for the lightest eigenstate. The dashed lines
	enclose the region where the observed dark matter density would be reproduced (within $2\sigma$'s)
	if the Majoron were produced thermally.}
	\label{fig:maj_v1_mj}
\end{figure}

Using Eq. (\ref{eq:Jtonu}), the bounds (\ref{eq:gammanu})-(\ref{eq:gammanu2}) on the decay rate can be expressed in terms of the underlying
particle physics parameter as \cite{Lattanzi:2013uza,Audren:2014bca}.
\begin{equation}
\frac{v_1}{\mathrm{GeV}} \ge  1.1 \times 10^{8} \left(\frac{m_J}{\mathrm{keV}}\right)^{1/2}
\left(\frac{\sum_i (m^\nu_i)^2}{\mathrm{eV}^2}\right)^{1/2} \quad\mathrm{WMAP9}\,, 
\label{eq:v1bound}
\end{equation}
\begin{equation}
\frac{v_1}{\mathrm{GeV}} \ge  3.7 \times 10^{8} \left(\frac{m_J}{\mathrm{keV}}\right)^{1/2}
\left(\frac{\sum_i (m^\nu_i)^2}{\mathrm{eV}^2}\right)^{1/2} \quad\mathrm{Planck2013+others}\,.
\label{eq:v1bound2}
\end{equation}

These limits are shown in Fig. \ref{fig:maj_v1_mj}. 

\begin{figure}[tb]
	\centering
	\includegraphics[width=0.5\columnwidth]{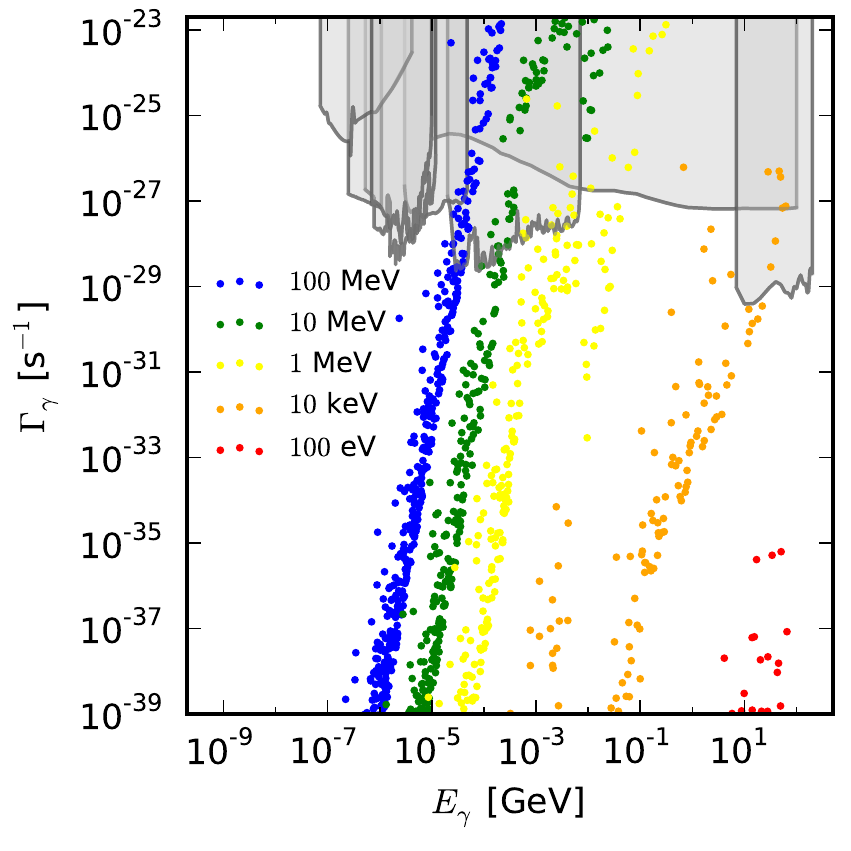}
	\caption{Constraints on the dark matter decay rate into two photons from a large variaty of astrophysical observations. Points show predictions for majoron models for different values of the triplet vev $v_3.$ See \cite{Lattanzi:2013uza} for more details.}
	\label{fig:maj_X}
\end{figure}

For masses in the range $100\,\mathrm{eV} \lesssim m_J \lesssim 100\,\mathrm{GeV}$, the sub-leading majoron decay to two photons of Eq. (\ref{eqJ:gg}), arising in the more
general see-saw scenario, can also be constrained through a number
of X- and $\gamma$-ray astrophysical observations \cite{bazzocchi:2008fh,Lattanzi:2013uza}. This decay mode provides an interesting route to probe majoron dark matter.
In particular, the $J\to\gamma\gamma$ constraints from line emission searches already
exclude part of the parameter space for models with $v_3$ larger than a few MeVs \cite{Lattanzi:2013uza}. A summary of current astrophysical constraints and theoretical predictions is shown in Fig.~\ref{fig:maj_X}. Recently, majoron decay into two photons was also proposed~\cite{Queiroz:2014yna}
as a plausible explanation of the line signal at 3.5~keV reported in Ref.~\cite{Bulbul:2014sua} (see next section for more details). 

\section{Sterile neutrino as dark matter}
\label{sec:sterile}

As discussed in Section~\ref{sec:mass_gen} sterile neutrinos are responsable for the generation of neutrino masses in the context of the \emph{type-I} seesaw mechanism. These particles interact with matter via the mass mixing with the active neutrino states, which can be parametrized in terms of the active-sterile mixing angle $\theta\ll 1$:
\begin{equation}
	\theta^2 \simeq \sum \left|\frac{m_D}{M_R}\right|^2 \ ,
\end{equation}

where the sum runs over all the sterile neutrino species. The mixing angle depends on the number of sterile neutrino states present in the model and
the relative size and texture of the Dirac and Majorana mass matrices. In general terms, this quantity comes from the diagonalization mass matrix described in Eq.~(\ref{eq:mass_matrix}).\\
The size of the mixing, usually expressed as $\sin^2{(2 \theta)}$, and the sterile neutrino mass $m_{\rm{s}}$  characterize the phenomenology of the sterile neutrino, i.e. its lifetime, decay modes and the production rate in the early Universe.
If the mixing angle is sufficiently small, sterile neutrinos are long lived and they can act as decaying dark matter candidates.
In order to account for the neutrino masses and explain dark matter at least three sterile neutrinos should be introduced. In the minimal realization, known as \emph{Minimal Neutrino Standard Model} ($\nu$MSM) \cite{2005PhLB..620...17A, 2005PhLB..631..151A}, the lightest sterile state acts as a dark matter candidate while neutrino masses and oscillations mainly depend on the other two heavier states. 
Interestingly, the decays of heavy sterile neutrinos in the early Universe could also explain the observed baryon-antibaryon asymmetry~\cite{2013PhRvL.110f1801C}.

\begin{figure}[tb]
	\centering
	\includegraphics[width=0.9\columnwidth]{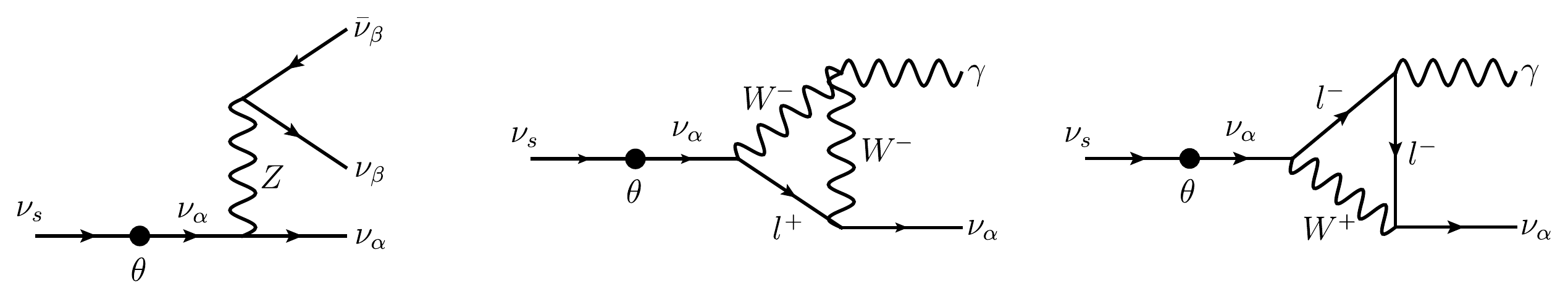}
	\caption{Sterile neutrino decay modes.} 
	\label{fig:nu_diags}
\end{figure}

\subsection{Lifetime and decay modes}
%
%

\begin{figure}[tb]
\centering
\includegraphics[width=0.9\columnwidth]{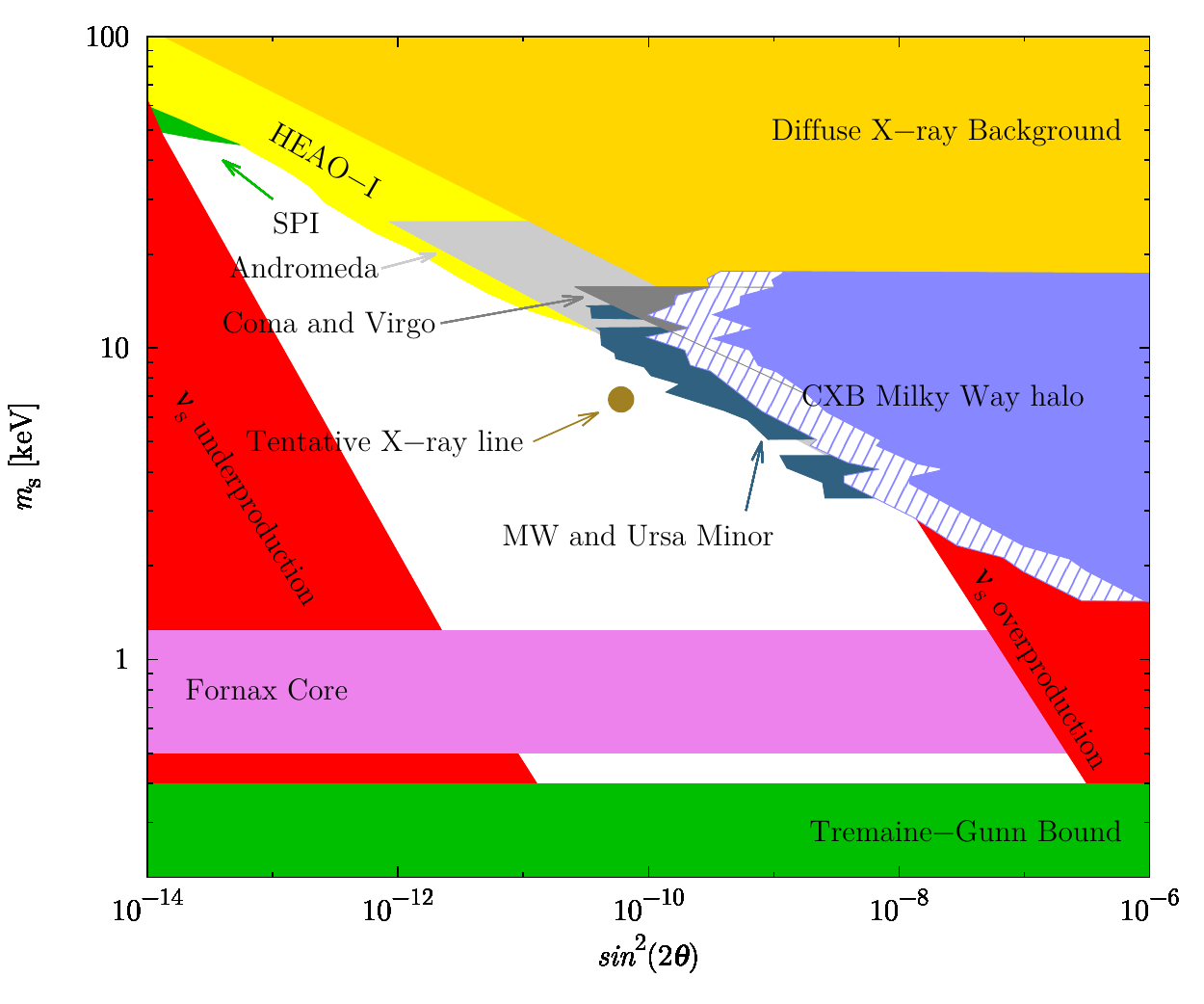}
\caption{
\label{fig:sterile_bounds}
Summary of constraints on sterile neutrino dark matter in the plane the sterile neutrino mass $(m_{\rm s})$ versus the mixing angle $sin^2{(2\theta)}$. The interpretation of the 3.5 keV X-ray line in terms of sterile neutrinos is shown in brown~\cite{Bulbul:2014sua}. In the red regions sterile neutrinos are either under-abundant or over-abundant~\cite{2009JCAP...03..005B}. The Green region corresponds to the Tremaine-Gunn bound~\cite{1979PhRvL..42..407T}. The pink one shows the constraint obtained from observations of the stellar velocity dispersion in the Fornax dwarf galaxy~\cite{2006ApJ...652..306S}. The rest of the regions are based on analysis of different celestial bodies in X-rays: Milky Way and Ursa minor (dark blue)~\cite{2007A&A...471...51B}, Andromeda (M31)~\cite{2006PhRvD..74c3009W}, Coma an Virgo clusters~\cite{2006PhRvD..74j3506B}, cosmic X-ray background assuming Milky Way halo models (light blue including hatching region~\cite{2006MNRAS.370..213B}), Milky Way SPI-INTEGRAL exclusion~\cite{2008MNRAS.387.1345B}, Milky Way HEAO-I exclusion~\cite{2006PhRvL..97z1302B}, and the diffuse X-ray background~\cite{2006MNRAS.370..213B}.}
\end{figure}

Sterile neutrino dark matter decays into SM particles with the diagrams shown in Fig.~\ref{fig:nu_diags}.
The lifetime can be expressed as~\cite{2012JCAP...10..014R,2012JHEP...06..100R}:
\begin{equation}
(\tau_{\rm s})^{-1} = \Gamma_{\rm s} \simeq \frac{G^2_{F} m_{\rm s}^5}{96 \pi^2} \times \mathcal{O}(\theta^2) \simeq 1.36 \times 10^{-29} \, {{\rm s}^{-1}} \left(\frac{\sin^2{(2\theta)}}{10^{-7}}\right) \left(\frac{m_{\rm s}}{1 \, {\rm keV}}\right)^5 ,
\end{equation}
where $G_F$ is the Fermi constant.\\ 

Small mixing angles are needed in order to push the lifetime beyond the age of the Universe. 
The loop-induced decays into $\gamma \nu $ induce an almost monochromatic photon emission which can be searched for with astrophysical observations. This decay mode provides therefore a way to test the sterile neutrino scenario, in analogy with the $\gamma\gamma$ signal for the majoron.
For sterile neutrino masses in the keV range, the $\gamma \nu$ line falls in the X-ray band. Searches of these signals have been performed by different X-ray observatories, focussing on a large number of targets. 
The corresponding constraints on the sterile neutrino parameter space are summarized in Fig.~\ref{fig:sterile_bounds}.\\

Recently it has been reported an identified and weak line signal at 3.5~keV~\cite{Bulbul:2014sua}, that can be interpreted as monochromatic emission from a decaying dark matter candidate with a mass of 7.1~keV. The analysis has been performed on stacked observations of 73 galaxy clusters with the XMM-Newton telescope, and the Perseus cluster with the Chandra X-ray Space Telescope. A consistent signal has been identified by another group analyzing the X-ray spectra of the Andromeda galaxy, of the Perseus cluster and of the Milky Way center~\cite{Boyarsky:2014jta,Boyarsky:2014ska}.

The statistical significance for the line emission is moderate and subject to modeling uncertainties. At present,
an explanation in terms of an atomic line emission can not be excluded, and it is under debate within the scientific community~\cite{Jeltema:2014qfa,Bulbul:2014ala,Boyarsky:2014paa}. The DM origin of the line has also been challenged by other X-ray observations of galaxies~\cite{2014arXiv1405.7943R,Malyshev:2014xqa,Anderson:2014tza}.
Summarizing, it is not yet possible to draw any firm conclusion on the origin of the 3.5~keV line.
Further observations and analyses are necessary to confirm the excess, and eventually to reject any standard astrophysical interpretation of this emission.
Interestingly, in case this signal would be confirmed, sterile neutrino could offer a viable explanation, as shown in Fig.~\ref{fig:sterile_bounds}. The signal can also be interpreted in terms of majoron decays into 2 photons, as proposed in~\cite{Queiroz:2014yna}. 
\subsection{Production mechanism in the Early Universe}
Sterile neutrinos are produced in the early Universe through oscillations with the active neutrino states, the so-called Dodelson-Widrow mechanism~\cite{1994PhRvL..72...17D}. As discussed in the previous section, astrophysical observations and the requirement of a sufficient long lifetime force the active-sterile mixing angle to be small. For these small mixings, sterile neutrino dark matter does not reach a thermal equilibrium with the plasma in the Universe. These particles are produced with a non thermal velocity spectrum and highly relativistic. They become nonrelativistic during the radiation-dominated epoch, behaving therefore as WDM. 
Their free-streaming length determines the scale below which the matter power-spectrum is damped compared to the CDM scenario. This suppression can be probed using the Lyman-$\alpha$ observations~\cite{1997ApJ...486..599H}. Bounds on the free-streaming length from Lyman-$\alpha$ can be traduced in lower limits on the mass of the sterile neutrino of the order of few keV~\cite{Viel:2013fqw,Boyarsky:2008xj}. Further lower bounds on the sterile neutrino mass are obtained from the phase-space dark matter density of dwarf-galaxies (the Tremaine-Gunn bound~\cite{1979PhRvL..42..407T}) and their density profiles, e.g.~\cite{2006ApJ...652..306S}.
Assuming the standard Dodelson-Widrow scenario, these constraints, combined with the bounds on X-rays observations, are in tension with the values of sterile neutrino masses and mixing angles for which the correct dark matter abundance is obtained~\cite{Horiuchi:2013noa}.

However, the sterile neutrino production is modified in presence of a lepton asymmetry in the Universe significantly larger than the baryonic one\footnote{We also remind that the sterile neutrino abundance can be diluted by a significant production of entropy in the early Universe~\cite{Bezrukov:2009th,Nemevsek:2012cd,King:2012wg}. Moreover sterile neutrinos might also be produced by decays of heavier particles, e.g.~\cite{Shaposhnikov:2006xi,Kusenko:2006rh,Merle:2013wta}.}. In this case, sterile neutrinos are produced via resonant neutrino oscillations, a mechanism known as \emph{resonant production} \cite{Shi:1998km}.
In this scenario sterile neutrinos can have the correct abundance in a much larger portion of the parameter space, namely  the region between the red areas in Fig.~\ref{fig:sterile_bounds}. In a specific point of this plane the correct sterile neutrino abundance is obtained for a proper value of the lepton asymmetry.
Moreover, in the case of resonant production the sterile neutrino velocity distribution is modified with respect to the non-resonant scenario, and the bounds from Lyman-$\alpha$ observations become less stringent~\cite{Boyarsky:2008mt}.

Summarizing, the sterile neutrino with a mass in the keV range is a viable decaying dark matter candidate. The constraints on its parameters discussed in these sections are shown in Fig.~\ref{fig:sterile_bounds}. 

\section{Flavour symmetries and dark matter}
\label{sec:flavour}

A fundamental property of dark matter is its stability over cosmological times. Naively, its lifetime should be larger than the age of the Universe,
$\sim 10^{18}$ s. Astrophysical observations pushes even further this bound. For instance, the observed photon background and measurements of cosmic-rays fluxes constrain the lifetime of dark matter to be larger than $10^{26}-10^{27}$ s for dark matter masses in the range $\sim 10-1000$ GeV (see \cite{Cirelli:2012ut} and references therein).

This apparent stability could be the result of the the smallness of the couplings involved in the dark matter decays. This is the case, for example, of the Majoron and sterile neutrino, discussed in the previous sections, or the gravitino in R-parity violating SUSY models \cite{Takayama:2000uz}.
Another option is that dark matter is stabilized by some symmetry, the most simple solution is a $Z_2$ parity. In many dark matter models the stabilization symmetry is imposed by hand, without any explanation for its origin. However, its existence could be intimately related to the structure of the theory and it could also play a central role in determining the interactions of dark matter. Therefore it would be theoretically more appealing to motivate the presence of such a symmetry in dark matter models. 
Different mechanisms have been proposed for this purpose,  for instance gauge symmetries \cite{Frigerio:2009wf,Kadastik:2009dj,Batell:2010bp} (e.g. the R-parity in SUSY models could arise from a $U(1)_{B-L}$ \cite{Martin:1992mq}), global symmetries and accidental symmetries \cite{Cirelli:2005uq,Hambye:2008bq} (see \cite{Hambye:2010zb} for a review on the subject).
Interestingly, the stability of dark matter could also originate from a discrete flavor symmetry \footnote{The dark matter stability can be connected to flavor physics also in the context of Minimal Flavor Violation \cite{Batell:2011tc,Kile:2013ola,Lopez-Honorez:2013wla}.}.

Non abelian discrete flavor symmetries have been extensively studied in order to explain the pattern of neutrino mixing \cite{Morisi:2012fg,Altarelli:2010gt} \footnote{The connection between flavor symmetries and sterile neutrino dark matter models is discussed, for instance, in \cite{Shaposhnikov:2006nn,Merle:2013gea}}. These symmetries and their breaking patterns define the structure of the leptons (and possibly quark) mass matrix and in turns their mixing angles.
In reference~\cite{Hirsch:2010ru} it has been proposed that the breaking of a flavor symmetry could leave a remnant symmetry stabilizying the dark matter.
A concrete example is presented, based on a $A_4$ flavor symmetry. $A_4$ is the group of even permutations of four objects and contains 12 elements. 
The generators of the group, in the 3 dimensional unitarity representations are:

\begin{equation}\label{eq:ST}
S=\left(
\begin{array}{ccc}
1&0&0\\
0&-1&0\\
0&0&-1\\
\end{array}
\right)\,;\quad
T=\left(
\begin{array}{ccc}
0&1&0\\
0&0&1\\
1&0&0\\
\end{array}
\right)\,.
\end{equation}

In this model the scalar potential of the SM is supplemented by three $SU(2)$ scalar doublets, $\eta=(\eta_1,\eta_2,\eta_3)$ which transform as an $A_4$ triplet, while the SM scalar $SU(2)$ doublet, $H,$ is an $A_4$ singlet. The minimization of the scalar potential gives the symmetry breaking pattern $\langle H \rangle=v_H$ and $\langle \eta \rangle =(v_{\eta},0,0).$
This breaks $A_4$ into a residual $Z_2$ symmetry generated by S and acting on the $A_4$ triplet $\eta$ as

\begin{equation}S \eta =\left(
\begin{array}{ccc}
1&0&0\\
0&-1&0\\
0&0&-1\\
\end{array}
\right)\left(\begin{array}{c}\eta_1 \\\eta_2\\\eta_3\end{array}\right)=\left(\begin{array}{c}\eta_1 \\-\eta_2\\-\eta_3\end{array}\right).\end{equation}
Since in this scenario the SM particles are even under this $Z_2$ parity, the lightest $Z_2$ odd particle of the model is automatically stable and can play the role of dark matter.

In this scheme neutrino masses are generated through a type-I seesaw mechanism introducing four heavy right handed neutrinos. With a suitable assignement of the $A_4$ charges of the RH neutrinos, the lepton doublets and the RH charged leptons, this model can accomodate the solar and atmospheric mixing angles and mass differences\footnote{The original model was constructed in such a way to give a vanishing reactor mixing angle $\theta_{13}.$ While this value was acceptable when the model was proposed, a non-vanishing $\theta_{13}$ has now been measured \cite{Adamson:2011qu,Abe:2013xua,Abe:2011fz,An:2012eh,Ahn:2012nd}. Recently, it  has been proposed that including radiative corrections, the model can reproduce the correct value of $\theta_{13}$~\cite{Hamada:2014xha}.}.
During the last years, the connection between a possible flavour structure of the leptonic sector and the stability of the dark matter, has been investigated in other models, considering also different flavor symmetries~\cite{Boucenna:2012qb,Lavoura:2012cv}.
In short, models based on flavor symmetries can lead to different patterns of the neutrino mass matrix and give a remnant or accidental $Z_2$ stabilizying the dark matter (see also~\cite{Meloni:2010sk,deAdelhartToorop:2011ad,Meloni:2011cc,Eby:2011qa,Adulpravitchai:2011ei,Kajiyama:2011fe,Kajiyama:2011gu}).

The precise properties of the dark matter candidate depends on the details of the model under consideration.
In the examples presented in refs.~\cite{Hirsch:2010ru,Boucenna:2012qb,Lavoura:2012cv,Meloni:2010sk}, the dark matter candidate is identified with a neutral $Z_2 $-odd scalar particle, more precisely the lightest mass eigenstate of the $\eta_2-\eta_3$ scalar system presented above. Dark matter can communicate with the SM  via ``Higgs portal'' interactions, induced by terms in the scalar potential like for instance $\eta^{\dagger}\eta H^{\dagger}H.$ In presence of ``weak-scale'' couplings, this dark matter candidate has a mass close to electro-weak scale and acts as a Weakly Interacting Massive Particle (WIMP). The correct cosmological relic abundance can be achieved through the standard thermal freeze-out mechanism.
As typical in WIMPs models, there exist multiple strategies to detect  these particles, namely searches at colliders, with underground dark matter detectors (direct searches) and with astrophysical observations (indirect searches).
In particular, the phenomenology of the dark matter candidate proposed in ref.~\cite{Hirsch:2010ru} has been extensively studied in~\cite{Boucenna:2011tj}.
Direct detection constraints exclude large regions of the parameter space while current indirect detection searches are sensitive to low dark matter masses.

\section{Conclusions}

We have reviewed the possibility that dark matter is somewhat related to the origin of neutrino masses.
In particular, we have shown how see-saw models, already in their minimal implementation, could at the same time generate neutrino masses and provide viable dark matter candidates.  Two simple and theoretically motivated examples are the sterile neutrino and the majoron. In the sterile neutrino scenario, the dark matter is provided by the lightest of the singlet right-handed neutral fermions, which are fundamental ingredients
of the seesaw mechanism. In the majoron scenario, on the other hand, the dark matter is associated to the scalar field responsible
for the dynamical generation of the majorana mass term of the right-handed neutrinos.

We have shown that both the sterile neutrino and the majoron can act, in some regions of the parameter space of the respective models, as decaying dark matter candidates. Moreover, requiring that their cosmological abundance matches the observed dark matter density, as inferred by CMB and other observations, roughly single out the keV mass range for both particles.

Interestingly enough, late decays of these candidates in monochromatic keV photons give an handle to identify them with astrophysical X-rays observations.
Current searches, in combination with the bounds from structure formation, are testing some portions of the parameter space of these models, as we have reviewed in the previous sections. Prospects with future experiments are discussed in~\cite{Neronov:2013lqa,Boyarsky:2012rt}. We also
mention the recent proposal for a new fixed-target experiment at the CERN SPS accelerator
that will use decays of charm mesons to search for heavy neutral leptons \cite{Bonivento:2013jag}.

The generation of neutrino masses can be related to dark matter also in more complicated, non-minimal, models.
The masses and properties of these dark matter candidates can be very different from those of the  majoron or the sterile neutrino.
As a concrete example, we have discussed the case of a WIMP-like, stable dark matter candidate. In particular we have considered the possibility that the stability of the dark matter particles are connected with the existence of a discrete $A_4$ flavour symmetry of the neutrino sector.

\ack
This work is by the Spanish MINECO under grants FPA2011-22975 and MULTIDARK CSD2009-00064 (Consolider-Ingenio 2010 Programme), by Prometeo/2009/091 (Generalitat Valenciana), and by the EU ITN UNILHC PITN-GA-2009-237920. R.L. is supported by a Juan de la Cierva contract (MINECO). 
M.T. is supported  by the European Research Council ({\sc Erc}) under the EU Seventh Framework Programme (FP7/2007-2013) / {\sc Erc} Starting Grant (agreement n. 278234 - `{\sc NewDark}' project).
M.L. is supported by Ministero dell'Istruzione, dell'Universit\`a e
della Ricerca (MIUR) through the PRIN grant \emph{`Galactic and extragalactic polarized microwave
emission'} (contract number PRIN 2009XZ54H2-002). Most of this work was carried out
while M.L. was visiting the Instituto de F\'isica Corpuscular, whose hospitality is
kindly acknowledged, supported by the grant \emph{Giovani ricercatori} of the University of Ferrara,
financed through the funds \emph{Fondi 5x1000 Anno 2010} and \emph{Fondi Unicredit
2013}.
R.L. also acknowledges the participants of the workshop: \emph{What is Dark Matter?} carried at NORDITA for fruitful discussions.\\

\section*{References}

\def\apj{Astrophys.~J.}                       
\def\apjl{Astrophys.~J.~Lett.}                
\def\apjs{Astrophys.~J.~Suppl.~Ser.}          
\def\aap{Astron.~\&~Astrophys.}               
\def\aj{Astron.~J.}                           %
\def\araa{Ann.~Rev.~Astron.~Astrophys.}       %
\def\mnras{Mon.~Not.~R.~Astron.~Soc.}         %
\def\physrep{Phys.~Rept.}                     %
\def\jcap{J.~Cosmology~Astropart.~Phys.}      
\def\jhep{J.~High~Ener.~Phys.}                
\def\prl{Phys.~Rev.~Lett.}                    
\def\prd{Phys.~Rev.~D}                        
\def\nphysa{Nucl.~Phys.~A}                    

\bibliographystyle{JHEP}

\bibliography{biblio}

\end{document}